\documentstyle[12pt]{article}
\newcommand{\address}[1]%
{\centerline{\small\it #1}}
\renewcommand{\title}[1]{\begin{center}%
{\large\bf #1}\end{center}\par\bigskip}
\renewcommand{\author}[1]{\centerline{#1}}
\renewcommand{\maketitle}{}
\newcommand{\pacs}[1]{}
\newcommand{\narrowtext}{\par\bigskip\bigskip}

\begin{document}
\title{Comment on 
``Extraction of work from 
a single thermal bath in the quantum regime''}
\author{Hal Tasaki$^*$}
\address{Department of Physics, Gakushuin University,
Mejiro, Toshima-ku, Tokyo 171, JAPAN}
\date{September 5, 2000}
\maketitle

\narrowtext
In a recent Letter \cite{AN}
Allahverdyan and Nieuwenhuizen
presented a series of calculations
on a quantum system coupled to a quantum
heat bath, and made a rather 
sensational conclusion
that {\em the second law of thermodynamics
may be violated}\/ as a
``consequence of quantum 
coherence in the 
presence of the slightly off-equilibrium 
nature of the bath.''

However, as we shall explain here,
one can use the standard result about
relative entropy to
prove {\em rigorously} that \cite{prev}
{\em the second law is never violated 
(and
a perpetual motion of the second kind 
can never be realized)
in quantum systems}
no matter how strong 
``quantum coherence'' is or 
no matter how far one goes from equilibrium.
The main purpose of the present Comment is 
not to make detailed objections to
\cite{AN}, but to remind the readers
that (under reasonable definitions which
are slightly different from those in 
\cite{AN}) the second law is never violated.

As in \cite{AN}, we consider a quantum system 
which consists of a subsystem 
(with Hilbert space 
\( {\cal H}_{\rm s} \))
and a bath
(with Hilbert space 
\( {\cal H}_{\rm b} \)).
The Hamiltonian of the subsystem 
\( H_{\rm s}(t) \)
depends on time
(where the time-dependence models 
operations by an outside agent),
that of the bath
\( H_{\rm b} \)
is time-independent, and that 
for interaction 
\( H_{\rm int}(t) \)
may be time-dependent.
The total Hamiltonian is
\( H(t)=H_{\rm s}(t)\otimes{\bf 1}+
{\bf 1}\otimes H_{\rm b} + 
H_{\rm int}(t) \).

Initially the subsystem is in an arbitrary
equilibrium state with density matrix
\( \rho^{\rm init}_{\rm s} \)
(on \( {\cal H}_{\rm s} \)),
and the bath is in the Gibbs state with
inverse temperature \( \beta \).
The density matrix for the whole system
is
\begin{equation}
	\rho^{\rm init}
	=
	\rho^{\rm init}_{\rm s}
	\otimes
	\frac{\exp[-\beta H_{\rm b}]}
	{Z(\beta)},
	\label{eq:rinit}
\end{equation}
where
\( Z(\beta)={\rm Tr}_{\rm b}
[\exp\{-\beta H_{\rm b}\}] \)
is the partition function for the bath.
(\( \rm Tr_{\rm s} \),
\( \rm Tr_{\rm b} \), and
\( \rm Tr \) stand for the traces over
\( {\cal H}_{\rm s} \),
\( {\cal H}_{\rm b} \),
and 
\( {\cal H}_{\rm s}
\otimes
{\cal H}_{\rm b} \), respectively.)

Let \( \rho^{\rm fin} \) be the 
density matrix at the final moment
obtained from the time evolution
according to \( H(t) \).
We make {\em no assumptions}\/ about the
nature of the time evolution or of the
final state \( \rho^{\rm fin} \).
{}From the well-known invariance of the 
von Neumann entropy and the 
nonnegativity of relative entropy
\cite{W},
we have
\begin{equation}
	-{\rm Tr}[\rho^{\rm init}
	\log\rho^{\rm init}]
	=
	-{\rm Tr}[\rho^{\rm fin}
	\log\rho^{\rm fin}]
	\le
	-{\rm Tr}[\rho^{\rm fin}
	\log\rho^{\rm ref}],
	\label{eq:ineq}
\end{equation}
where 
\( \rho^{\rm ref} \)
is an arbitrary density matrix.
{}From (\ref{eq:rinit}) one finds that
\( -{\rm Tr}[\rho^{\rm init}
\log\rho^{\rm init}]
=
S_{\rm vN}[\rho^{\rm init}_{\rm s}]
+\beta\langle H_{\rm b}\rangle_{\rm init} 
+\log Z(\beta) \),
where
\( S_{\rm vN}[\rho_{\rm s}]
=-{\rm Tr}_{\rm s}
[\rho_{\rm s}\log\rho_{\rm s}] \)
is the von Neumann entropy of the subsystem.
If we choose the reference state as 
\begin{equation}
	\rho^{\rm ref}
	=
	\rho^{\rm fin}_{\rm s}
	\otimes
	\frac{\exp[-\beta H_{\rm b}]}
	{Z(\beta)},
	\label{eq:rref}
\end{equation}
where 
\( \rho^{\rm fin}_{\rm s}
=
{\rm Tr}_{\rm b}[\rho^{\rm fin}] \),
we have
\( -{\rm Tr}[\rho^{\rm fin}
\log\rho^{\rm ref}]
=
S_{\rm vN}[\rho^{\rm fin}_{\rm s}]
+\beta\langle H_{\rm b}\rangle_{\rm fin} 
+\log Z(\beta) \).
Therefore (\ref{eq:ineq}) becomes
\begin{equation}
	S_{\rm vN}[\rho^{\rm fin}_{\rm s}]
	-
	S_{\rm vN}[\rho^{\rm init}_{\rm s}]
	\ge
	\beta\left(
	\langle H_{\rm b}\rangle_{\rm init}
	-
	\langle H_{\rm b}\rangle_{\rm fin}
	\right).
	\label{eq:Clausius}
\end{equation}
Since 
\( \langle H_{\rm b}\rangle_{\rm init}
-
\langle H_{\rm b}\rangle_{\rm fin} \)
is the energy transferred from the bath
to the rest of the system
(described by \( H_{\rm s} \) plus
\( H_{\rm int} \)),
it can be unambiguously interpreted
as the {\em heat}\/
that flowed out of the bath
during the process \cite{heat}.
We have thus confirmed the validity of the
second law
(in the form of the Clausius inequality)
rigorously \cite{TD}.
We note in passing that (\ref{eq:Clausius})
{\em rules out the possibility of any 
perpetual motion
of the second kind,
including those operating
in nonequilibrium states with strong quantum
coherence}\/ \cite{pm}.

\bigskip
It is a pleasure to thank
Elliott Lieb,
Keiji Saito,
Shin-ichi Sasa,
Jakob Ynvgason,
and
Satoshi Yukawa
for 
valuable discussions.
I also thank
Armen Allahverdyan
and
Theo Nieuwenhuizen
for useful comments.


\begin{thebibliography}{10}
\bibitem{heat}
If one is worried about the treatment of
\( H_{\rm int} \),
one can set \( H_{\rm int}(t) \) to be
vanishing at the initial and the 
final moments (while it can be large 
in between them).
Our definition is different from that in \cite{AN},
but we believe that ours
corresponds to the energy 
``extracted from the bath'' in the form of heat.
\bibitem{TD}
\( S_{\rm vN}[\rho_{\rm s}] \)
may not in general coincide with 
the thermodynamic entropy which is defined
(for example) as
\( S_{\rm TD}[\rho_{\rm s}]
=
\sup_{\rho}
\{S_{\rm vN}[\rho]\,|\,
{\rm Tr}_{\rm s}[\rho H_{\rm s}]
=
{\rm Tr}_{\rm s}[\rho_{\rm s}H_{\rm s}]
\} \).
This does not alter our conclusion
since we can guarantee
\( 
S_{\rm TD}[\rho_{\rm s}^{\rm init}]
=
S_{\rm vN}[\rho_{\rm s}^{\rm init}]
\)
by choosing
\( \rho_{\rm s}^{\rm init} \)
(for example) as a Gibbs state,
and can use
\( 
S_{\rm TD}[\rho_{\rm s}^{\rm fin}]
\ge
S_{\rm vN}[\rho_{\rm s}^{\rm fin}]
\)
for the final state to prove
the inequality (\ref{eq:Clausius})
with \( S_{\rm vN} \)
replaced by
\( S_{\rm TD} \).
For details, see
H. Tasaki, in preparation.
\bibitem{pm}
To see this, suppose that one starts from 
(equilibrium and non-coherent)
\( \rho^{\rm init} \),
brings the system into a 
(perhaps non-equilibrium and coherent)
cycle which violates Kelvin's principle,
repeat the cycle for \( N \) times, 
and then stop it.
Then the left-hand side of (\ref{eq:Clausius})
is independent of \( N \) while the
right-hand side consists of 
a constant
plus a positive term proportional to 
\( N \).
This leads to a contradiction for
large enough \( N \) (which can be made
arbitrarily large by making the bath large).
In \cite{AN} it is indeed noted that their 
``perpetuum mobile'' may be repeated
for a finite number of cycles, which, in our opinion,
simply means that it should not be called a 
``perpetuum mobile.''
\end{thebibliography}
\end{document}